\documentclass[doublecol]{epl2}
\usepackage{amsmath}
\usepackage{graphicx}
\usepackage{stfloats}

\title{Ultra accurate collaborative information filtering via directed user similarity}
\shorttitle{Ultra accurate collaborative information filtering via directed user similarity} 

\author{Q. GUO \and W.-J. SONG \and J.-G. LIU\thanks{E-mail:liujg004@ustc.edu.cn}}
\shortauthor{Q. GUO {\it et al}}

\institute{
Research Center of Complex Systems Science, University of
Shanghai for Science and Technology, Shanghai 200093, PR China}


\abstract{A key challenge of the collaborative filtering (CF) information filtering is how to obtain the reliable and accurate results with the help of peers' recommendation. Since the similarities from small-degree users to large-degree users would be larger than the ones opposite direction, the large-degree users' selections are recommended extensively by the traditional second-order CF algorithms. By considering the users' similarity direction and the second-order correlations to depress the influence of mainstream preferences, we present the directed second-order CF (HDCF) algorithm specifically to address the challenge of accuracy and diversity of the CF algorithm. The numerical results for two benchmark data sets, MovieLens and Netflix, show that the accuracy of the new algorithm outperforms the state-of-the-art CF algorithms. Comparing with the CF algorithm based on random-walks proposed in the Ref.7, the average ranking score could reach 0.0767 and 0.0402, which is enhanced by 27.3\% and 19.1\% for MovieLens and Netflix respectively. In addition, the diversity, precision and recall are also enhanced greatly. Without relying on any context-specific information, tuning the similarity direction of CF algorithms could obtain accurate and diverse recommendations. This work suggests that the user similarity direction is an important factor to improve the personalized recommendation performance.}

\begin{document}

\maketitle

\section{Introduction}
With the rocketing development of the Internet, we are confronted with the problem of information overload \cite{1,2}. In order to break through this dilemma, various recommender algorithms \cite{3,4,5,6,7,8,9}, which attempts to predict users' interests by analyzing their historical activities, have been proposed. So far, the collaborative filtering (CF) algorithm \cite{10,11} has been one of the successful recommendation algorithms, which is designed based on the assumption that users with similar preferences will rate similar objects. When predicting the potential interests of a given user, the CF algorithm firstly identifies the neighborhood of each user by calculating similarities between all pairs of users, and then makes recommendations based on the neighbors' selections. It is well known that the most important ingredient in determining the performance of the CF algorithm is how to precisely define the similarities between each pair of users \cite{12,13}. Based on the user-object bipartite network, the cosine similarity \cite{14} is the most widely used index to quantify the proximity of users' tastes. In addition, Sarwal {\it et al.} \cite{15} proposed the item-based CF algorithm by comparing different items. Deshpande and Karypis \cite{16} proposed the item-based top-$N$ CF algorithm, in which items were ranked according to the frequency of appearing in the set of similar items and the top-$N$ ranked items were returned. Luo {\it et al.} \cite{17} introduced the concepts of local and global user similarity based on surprisal-based vector similarity
and the concepts of maximum distance in graph theory.

Recently, some physical dynamics, such as random walks \cite{6,7} and heat conduction \cite{18}, have found their applications in user or item similarity measurement to generate recommendation algorithms. Liu {\it et al.} \cite{7} embedded the random-walks process into the CF algorithm to calculate the user similarity and found that the random-walk-based CF algorithm had remarkable accuracy. By taking into account the second-order correlation of the objects and users, Zhou {\it et al} \cite{19} proposed improved CF algorithms by depressing the influence of mainstream preferences. The simulation results show that both accuracy and diversity of the improved CF algorithms could be enhanced greatly. By tuning the similarity from neighbors to the target user, Liu {\it et al.} \cite{20} analyzed the relationship between the similarity direction and the performance of the CF algorithm and found that emphasizing the small-degree users' recommendation powers could not only accurately identify users' interests, but also increase the algorithmic capability of finding diverse objects. Inspired by the idea that both second-order user correlations and the similarity direction affect the accuracy and diversity of the CF algorithms, we investigate the effect of the similarity direction on the second-order CF algorithm. Based on the statistical properties of the user correlation network and the similarity direction effect, we present a modified CF algorithm, namely the directed second-order CF (HDCF) algorithm. The experimental results on the data sets Movielens and Netflix show that only by changing the direction of the similarity obtained by the second-order user correlation, the accuracy greatly outperforms state-of-the-art CF methods, which suggests that the similarity direction is definitely a significant factor for second-order information filtering.

\begin{center}
\begin{figure}[t]
\center\scalebox{0.75}[0.75]{\includegraphics{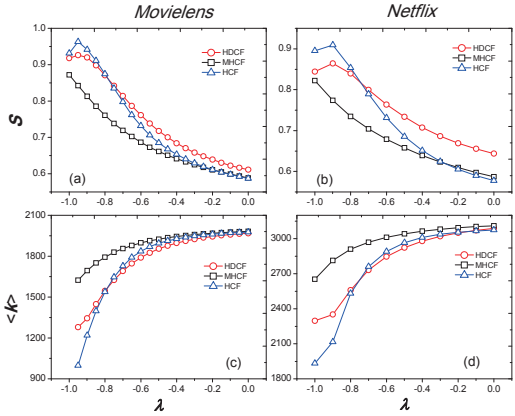}}
\caption{(Color Online) The diversity $S$ and the average object
degrees $\langle k\rangle$ vs. $\lambda$ of the HDCF, HCF and MHCF
algorithms for Movielens and Netflix data sets when recommendation
list equals to $L=10$. (a)-(b) exhibit the diversity $S$ vs.
$\lambda$, and (c)-(d) show the average object degrees $\langle
k\rangle$ vs. $\lambda$. At the optimal cases, both diversity $S$
and popularity $\langle k\rangle$ of HDCF are much better than the
ones of MHCF algorithms. All the data sets points are averaged over
ten independent runs with different data set divisions.}
\label{Fig2}
\end{figure}
\end{center}

\begin{center}
\begin{figure}[h]
\center\scalebox{0.75}[0.75]{\includegraphics{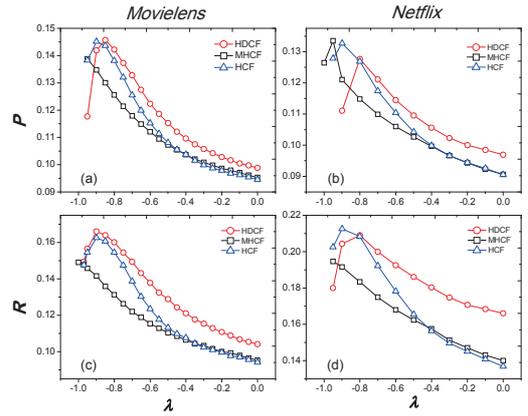}}
\caption{(Color Online) Precision $P$ and Recall $R$ of the HDCF,
HCF and MHCF algorithms for Movielens and Netflix data sets when the
recommendation list length $L=10$. (a)-(b) exhibit the precision $P$
vs. $\lambda$, and (c)-(d) report recall $R$ vs. $\lambda$. At the
optimal cases, both $P$ and $R$ of HDCF for Movelens and Netflix are
larger than the ones of MHCF algorithm. All the data points are
averaged over ten independent runs with different data set
divisions.} \label{Fig3}
\end{figure}
\end{center}

\section{Related works}
\subsection{Bipartite network and the standard CF algorithm}

An information filtering system could be characterized by a user-object bipartite network which consists of a set of user nodes denoted as $U=\{u_{1},u_{2},\ldots,u_{n}\}$, the object nodes as $O=\{o_{1},o_{2},\ldots,o_{m}\}$ and the links between these two sets, which are indicated by $E=\{e_{1},e_{2},\ldots,e_{p}\}$. The bipartite network containing $n$ users and $m$ objects can be represented by an adjacent matrix $A=\{a_{ij}\}\in R^{m,n}$, where $a_{ij}=1$  if user $u_{j}$ collects object $o_{i}$  and $a_{ij}=0$ otherwise.

Inspired by the random-walks process presented by Zhou {\it et al.} \cite{6}, Liu {\it et al.} \cite{7} proposed a CF algorithm based on the random-walks (denoted by CF). It is assumed that a certain amount of resource is associated with each user, and the weight $s_{ij}$
represents the proportion of the resource user $u_j$ would like to distribute to user $u_i$. The weight $s_{ij}$ representing the fraction of initial resource user $u_j$ ultimately gives to user $u_i$ can be defined as
\begin{equation}\label{equation1}
s_{ij}=\frac{1}{k_{u_j}}\sum_{l=1}^m\frac{a_{li}a_{lj}}{k_{o_l}},
\end{equation}
where $k_{u_j}=\sum\nolimits_{\alpha=1}^{m}a_{\alpha j}$ and $k_{o_l}=\sum\nolimits _{i=1}^{n}a_{li}$ denote the degrees of user $u_j$ and object $o_l$. For the user-object pair $(i,\alpha)$, if user $u_i$ has not collected object $\alpha$ yet (i.e., $a_{\alpha i}=0$), the predicted score $v_{i \alpha}$, is given as
\begin{equation}\label{equation2}
v_{i\alpha}=\frac{\sum_{j=1}^n s_{ji}a_{\alpha
j}}{\sum_{j=1}^ns_{ji}},
\end{equation}
where $s_{ji}$ indicates the similarity from the target user $u_i$ to its neighbor $u_j$. To the target user $u_i$, when the predicted scores $v_{i\alpha}$ among all the objects he/her has not collected are calculated, all these scores will be sorted in descending order, and finally those on the top will be recommended.

\subsection{Second-order user correlation effect analysis}
\begin{figure*}
\begin{equation}
\begin{array}{ccl}
h_{ij} & = & s_{ij} + \lambda \sum_u{s_{iu}s_{uj}},\\[5pt]
       & = & \frac{1}{k_{j}}\sum_{o=1}^m\frac{a_{o i}a_{o j}}{k_{o}}
       + \lambda
       \sum_{u=1}^{n}\Big(\frac{1}{k_{u}}\sum_{o=1}^m\frac{a_{oi}a_{ou}}{k_{o}}\Big)\Big(\frac{1}{k_{j}}\sum_{o=1}^m\frac{a_{ou}a_{oj}}{k_{o}}\Big)\\
  & = & \frac{1}{k_{j}}\Big\{\sum_{o=1}^m\frac{a_{o i}a_{o j}}{k_{o}}
       + \lambda
       \sum_{u=1}^{n}\Big(\frac{1}{k_{u}}\sum_{o_1=1}^m\sum_{o_2=1}^m\frac{a_{o_1i}a_{o_1u}}{k_{o_1}}\frac{a_{o_2u}a_{o_2j}}{k_{o_2}}\Big)\Big\}.\\
\end{array}
\end{equation}
\end{figure*}

The correlation between two users is a reflection of their similar tastes or preferences, therefore, for two arbitrary users, the specific interests should contribute more to the similarity measurement than the mainstream preferences \cite{add1}. In addition, two users sharing many mainstream preferences have high second-order similarity. Liu {\it et al.} \cite{8} proposed an effective method to depress the influence of mainstream preferences by considering the second-order similarity, where the similarity matrix is given by
\begin{equation}\label{equation8}
{\bf H}={\bf S}+\lambda {\bf S}^{2},
\end{equation}
and ${\bf H}=\{h_{ij}\}_{n,n}$ is the user similarity matrix obtained by taking into account the second-order user correlation. Here the value range of the parameter $\lambda$ is $(-1,0)$ in order to improve the algorithmic accuracy.

\subsection{The directed second-order CF algorithm}
\revision{To users $u_i$ and $u_j$, the similarity from user $u_j$ to user $u_i$, $h_{ij}$ representing the amount of initial resource $u_j$ evenly transferred to $u_i$ could be written as Eq.3. It is unlikely these quantities are exactly the same for each pair of users, therefore, $h_{ij}\neq h_{ji}$ in most cases. In addition, one has the following relationship
\begin{equation}
\frac{h_{ij}}{h_{ji}} = \frac{k_i}{k_j}.
\end{equation}
If $k_i>k_j$, then $h_{ij}>h_{ji}$ and vice versa. In other words, the similarities from small-degree users to large-degree users would be larger than the ones from the opposite direction. Since that the degree of most users in the real world is very small, which means that the large-degree users would frequently be identified as small-degree users' friends, the second-order CF algorithm would emphasize the large-degree users' recommendation powers, leading to the high similarity of most users' recommendation lists.

Furthermore, the second-order CF algorithm using the similarity from the target user to his neighbors, does not match up to the core idea of the standard CF algorithm. Therefore we could enhance the small-degree users' effects by reversing the user similarity direction from neighbors to the target user} and present the directed second-order CF (denoted by HDCF) algorithm, which could be described as follows
\begin{description}
\item[(i)] To user $i$, calculating the user similarity from all other users $\{h_{ij}\} (j=1,2,\cdots,n)$ in terms of the Eq.4;
\item[(ii)]
For each user $i$, calculating the predicted scores for the
uncollected objects,
\begin{equation}\label{equation9}
v_{i\alpha}=\frac{\sum_{j=1}^n h_{ij}a_{\alpha
j}}{\sum_{j=1}^n h_{ij}},
\end{equation}

\item[(iii)] Sorting the uncollected objects in the descending order of the predicted scores, and those objects on the top will be recommended.
\end{description}

\subsection{The maximum-similarity-based CF algorithm}

Actually the properties of the real datasets could also affect the algorithmic performance. In other words, although the performance of the HDCF algorithm outperforms the state-of-the-art CF algorithms, it may only happen in some certain datasets whose second-order similarities from neighbors to the target user are more effective than the ones in the reverse direction. Thus, we present a maximum-similarity-based CF (denoted by MHCF) algorithm to investigate the effect of second-order similarity magnitude, and the predicted score $v_{i\alpha}$ is given by
\begin{equation}\label{equation10}
v_{i\alpha}^{m}=\frac{\sum_{j=1}^n h^{\rm max}_{ij}a_{\alpha
j}}{\sum_{j=1}^nh^{\rm max}_{ij}},
\end{equation}
where $h^{\rm max}_{ij}$ is defined as the larger second-order similarity between users $u_i$ and $u_j$
\begin{equation}\label{equation11}
h^{\rm max}_{ij}=\max\{h_{ij}, h_{ji}\}.
\end{equation}

\section{Experimental results}
\subsection{Data description}

In this paper, we implement our experiments on two benchmark data sets, {\it Movielens} \footnote{http://www.Movielens.com} and {\it Netflix} \footnote{http://http.netflix.com}. Movielens consists of 6040 users and 3952 movies (objects). Netflix is a random sample of the original Netflix data set, which consists of 6000 movies and 10000 users, and 824802 ratings. The users of the data sets vote on movies with discrete ratings from 1 (i.e., worst) to 5 (i.e., best). Here we apply a coarse graining method \cite{6}: A movie is considered to be collected by a user only if the given rating is three or more. In this way, the Movielens data have 836478 edges, and the Netflix data have 701947 edges. Table 1 gives the basic statistical properties of the data sets. In order to evaluate the algorithmic performance, we randomly divide data set $E$ into two parts: $E=E^{T} \cup E^{P}$, where $E^T$ is the training set treated as known information, and the other one $E^P$ is the probe set, whose information is not allowed to be used for prediction. We treat 90\% percent of the ratings as the training set, and the remaining 10\% part compose the probe set. Then five different metrics are employed to test the algorithm performance, including average ranking score, popularity, diversity, precision and recall.

\begin{table}
\footnotesize
\caption{Basic statistical properties of the tested data sets.}
\begin{center}
\begin{tabular} {ccccc}
  \hline \hline
   Data Sets      &  Users   & Objects & Links & Sparsity \\ \hline
   MovieLens      &  6040    & 3952     & 836,478 &  $3.50\times 10^{-2}$\\
   Netflix        &  10,000  & 6,000  & 701,947    & $1.17\times 10^{-2}$ \\
   \hline \hline
    \end{tabular}
\end{center}
\end{table}

\subsection{Performance metrics}
1) {\bf Average ranking score} \cite{22}

Since the average ranking score doesn't rely on the length of recommendation list, we use it to measure the ability of the algorithm to produce a good uncollected object ranking list that matches the target user's preference. For an arbitrary user $u_i$, if in the training set the object $o_{\alpha}$ is not collected by user $u_i$, while the entry $(i,\alpha)$ is in the probe set, we use the rank of the object $o_\alpha$ in the recommendation list to evaluate accuracy. Therefore, the mean value of the positions, called {\it average ranking score} $\langle r \rangle$, averaged over all the entries in the probe set, can be used to evaluate the algorithmic accuracy
\begin{equation}\label{equation12}
\langle r\rangle=\frac{1}{n}\sum_{i=1}^n\Big(\frac{\sum_{(u_i,
o_\alpha)\in E^p}r_{i\alpha}}{q-k_{u_i}}\Big),
\end{equation}
where $E^{p}$ is the edge set existing in the probe set and $q$ is the number of objects in the probe set. The smaller the average ranking score, the higher the algorithmic accuracy, and vice versa.

2) {\bf Diversity} \cite{23}

Personalized recommendation algorithms should not only present accurate prediction, but also generate different recommendations to different users according to their specific tastes or habits. Therefore, besides accuracy, the {\it diversity} measured by the mean value of {\it Hamming distance} $S$, is taken into account to evaluate the strength of personalization. If the overlapped number of objects in $u_i$ and $u_j$ recommendation list $L$ is $Q_{ij}$, their Hamming
distance could be quantified as:
\begin{equation}\label{equation13}
S= 1-\langle Q_{ij}(L)\rangle/L,
\end{equation}
Generally speaking, a more personalized recommendation list should have larger Hamming distance to other lists. Accordingly, we adopt the mean value of Hamming distance $S=\langle H_{ij} \rangle$, averaged over all the user pairs, to measure the strength of the algorithmic diversity. The largest $S=1$ indicates recommendation to all users are completely different. While the smallest $S=0$ means all of the recommendations are exactly the same.

3) {\bf Popularity} \cite{22}

An accurate and diverse recommender system is expected to help users find the niche or unpopular objects that are hard for them to identify. Since there are countless channels which advertise the popular movies, such as the Internet, TV, newspaper, etc., uncovering the very specific preference, corresponding to unpopular objects, is much more significant than simply picking out what a users likes from the top of the list. The metric {\it popularity} is introduced
to quantify the ability of an algorithm to generate the unexpected recommendation lists, which is defined as the average collected times over all recommended objects:
\begin{equation}\label{equation14}
\langle k\rangle =
\frac{1}{n}\sum_i\Big(\frac{1}{L}\sum_{o_\alpha\in
O^L_i}k_{o_\alpha}\Big),
\end{equation}
where $o_{i}^{L}$ is user $i$'s recommendation list with length $L$. A smaller average degree $\langle k\rangle$, corresponding to less popular objects, is preferred since those small-degree objects are hard to be found by users themselves.

\begin{table*}
\scriptsize
\caption{Algorithmic performances for Movielens and Netflix
  data sets when $p=0.9$, including the average ranking score
  $\langle r\rangle$, and diversity $S$, popularity $\langle k\rangle$, precision $P$ and
  recall $R$
  corresponding to the length of recommendation list $L=10$.
  CF is the collaborative filtering algorithm based on
  random-walks proposed in the Ref.7; DCF is the directed random-walks-based
  CF algorithm whose similarity measurement is from neighbors
  to the target user ($\beta_{\rm opt}=3.2$ for Movielens and
  $\beta_{\rm opt}=2.0$ for Netlix) \cite{20}; HCF is an improved
  CF algorithm, in which the user similarity is based
  on the random-walks, and the second-order correlation is
  involved ($\lambda_{\rm opt}=-0.85$ for Movielens and
  $\lambda_{\rm opt}=-0.8$ for Netflix) \cite{25}; MHCF is the high-order
  CF algorithm whose similarity is defined as the larger one between
  two users ($\lambda_{\rm opt}=-0.95$ for Movielens and
  $\lambda_{\rm opt}=-0.9$ for Netflix); HDCF is
  the presented new algorithm in this paper. Each number is
  obtained by a averaging over ten runs of independently
  random division of training set and probe set.}
\begin{center}
\begin{tabular}{ccc|cccccccccccc}
\hline\hline
          &      & Algorithms && $\langle r\rangle$     & & $\langle k\rangle$   & & $S$  & & $P$  & & $R$ &\\ \hline
   &               &   CF     & & 0.1055    & & 1818  & & 0.7068 && 0.1025 && 0.1031&\\
     &             &   DCF    & & 0.0853    & & 1259  & & 0.9191 && 0.1053 && 0.1165&\\
 &{\it Movielens}  &   HCF    & & 0.0828    & & 1397  & & 0.9112 && 0.1436 && 0.1607&\\
       &           &   MHCF   & & 0.0791    & & 1623  & & 0.8424 && 0.1388 && 0.1456&\\
         &         &   HDCF   & &{\bf 0.0767}& & 1545 & & 0.8707 && 0.1422 & & 0.1602&\\
\hline
               &  &    CF     & & 0.050    & & 2813  & & 0.7001 && 0.0917 && 0.1365&\\
               &  &    DCF    & & 0.045    & & 2506  & & 0.8236 && 0.0967 && 0.1640&\\
 &{\it Netflix}   &    HCF    & & 0.0434    & & 2531  & & 0.8535 && 0.1269 && 0.2083&\\
               &  &    MHCF   & &{\bf 0.0402}&& 2814 && 0.7737 && 0.1210 && 0.1915&\\
                & &    HDCF   & & {\bf 0.0402}&& 2731 && 0.7997 && 0.1211 && 0.1998&
                \\ \hline\hline
\end{tabular}
\end{center}
\end{table*}

4) {\bf Precision and Recall} \cite{ad23,24}

Since real users are usually cared only about the top part of the recommendation list, a more practical method is to consider the number of a user's deleted links contained in the top-$L$ positions. Based on our concerns, we may take into account either how many of these top $L$ places are possessed by deleted links, or how many of user's deleted links have been recovered in this way. Thus, precision and recall can satisfy our requests. For an arbitrary user $u_i$, the precision and recall of the recommendation, $P_i(L)$ and $R_i(L)$, are defined as
\begin{equation}\label{equation15}
P_i(L)=\frac{d_i(L)}{L},\ \ R_i(L)=\frac{d_i(L)}{D_i}
\end{equation}
where $d_i(L)$ indicates the relevant objects in the top-$L$ positions of the recommendation list that collected by user $u_i$ in the probe set, and $D_i$ is the total number of user $u_i$'s relevant objects. Averaging the individual precision and recall over all users, we obtain the mean value {\it Precision} $P(L)$ and {\it Recall} $R(L)$ of the algorithm on one data set
\begin{equation}\label{equation16}
P(L)= \frac{1}{n}\sum_i\frac{d_i(L)}{L},\ \ R(L)=
\frac{1}{n}\sum_i\frac{d_i(L)}{D_i}.
\end{equation}
A larger precision corresponds to a better performance, and the larger recall corresponds to the better performance.

\subsection{Performance}

Table 2 shows the comparisons of five metrics among different CF algorithms at cutoff 10. The CF algorithm based on random-walks proposed by Liu {\it et al.} \cite{7} is denoted by CF, DCF represents the directed random-walks-based CF algorithm \cite{20}, HCF is the second-order-based CF algorithm \cite{25}, MHCF stands for the high-order-based CF algorithm whose similarity is defined as the larger one and the presented new algorithm in this paper considering the similarity direction and the second-order user correlation is called HDCF. Comparing with the HCF algorithm, the average ranking score $\langle r\rangle$ of the HDCF algorithm is reduced from 0.0828 to 0.0767 for Movielens and from 0.0434 to 0.0402 for Netflix, at the optimal values of $\lambda$. Apparently, by only taking into account the similarity direction of the second-order correlation without any other information, the accuracy of our algorithm outperforms the standard second-order CF algorithm, and even better than state-of-the-art CF algorithm. Clearly, enhancing the small-degree users' recommendation powers by changing the similarity direction could provide fairly accurate recommendation results.

Fig.1 represents the diversity $S$ and the popularity $\langle k\rangle$ as a function of $\lambda$ when the recommendation list $L=10$ for Movielens and Netflix, respectively. From Fig.1(a)-(b), one could find that, for different data sets, the diversity $S$ is negatively correlated with $\lambda$, which indicates that involving the directed second-order similarity makes the recommendation lists more diverse. Compared with the results of the standard CF algorithm, when recommendation list length $L=10$, for the Movielens data set, the diversity $S=0.8707$ at the optimal $\lambda_{\rm opt}=-0.8$, and for the Netflix, the diversity $S=0.7997$ at the optimal $\lambda_{\rm opt}=-0.7$, which is improved by 23.2\% and 14.2\% respectively. Fig.1(c)-(d) exhibit a positive relation between popularity $\langle k\rangle$ and $\lambda$, thus to depress the influence of mainstream preferences, we should give more opportunity to the less popular objects. Comparing with the results of standard CF algorithm corresponding, when $L=10$, the popularity $\langle k\rangle=1545$ for Movielens and $\langle k\rangle=2731$ for Netflix at their optimal values, which is reduced by 15\%, and by 2.9\% respectively. So the presented algorithm with negative $\lambda$ has the capability to provide more diverse recommendation lists and excavate unpopular objects.

Since users always pay more attention to the top of the recommendation lists, from Fig.2 we could see that the precision $P$ and recall $R$ are also very good. Comparing with the standard CF algorithm, when $L=10$ with the optimal parameter corresponding to the lowest ranking score ,the $P$ is roughly improved by 38.7\% and 32\%, and $R$ is approximately enhanced by 55.4\% and 46\% for Movielens and Netflix respectively.

In general, the specific common interests contribute more to the similarity measurement between two users than their mainstream preferences, and the directed similarity calculated by the random-walks process is reverse to the initial users' degrees. Thus, the standard CF algorithm may count repeatedly the attributes
of the popular objects which meet the tastes of most users and would assign more power for the larger-degree users to the small-degree users, which decreases the accuracy, increases the average object degrees and reduces the diversity. Our presented algorithm with negative $\lambda$ parameter and new user similarity direction could depress the influence of mainstream preferences and enhance the small-degree users' recommendation powers, which increases the
prediction accuracy greatly, as well as gives high chances to less popular objects and help users find diverse objects, leading to better algorithm performance.

\section{Conclusions and discussions}

The CF algorithms are one of the most successful information  filtering algorithms and have been extensively implemented to many online applications. In this paper, by considering the second-order user correlations and the similarity direction, we present a directed second-order CF algorithm by tuning the user similarity from neighbors to the target user to emphasize the recommendation powers of small-degree users. The experimental results for MovieLens and Netflix data sets show that the new algorithm could indeed generate a more favorable recommendation performance. Compared with the CF algorithm based on random-walks, the average ranking score could be improved to 0.0767 and 0.0402, which is enhanced by 27.3\% and 19.1\% for MovieLens and Netflix respectively. Additionally, the diversity, precision and recall are also enhanced greatly. The possible reasons for the pretty good performance of HDCF algorithm lie in the fact that second-order similarity could depress the influence of mainstream preferences on the target user, and tuning the similarity direction of the CF algorithm from the neighbors to the target user could enhance small-degree users' recommendation power to improve the accuracy and help users uncover less popular objects.

Since we only need to change the similarity direction of the tradition second-order CF algorithm without any other information, it is very easy for this algorithm to be used in real information filtering systems. How to better provide personalized recommendations by taking into account the mixing pattern \cite{26} of the network for diverse users is a long-standing challenge in modern information science. Any method to this issue may intensively change our society, economic and life in the near future. We believe our current work will enlighten readers in this interesting and promising direction.

\section{Acknowledgments}
We acknowledge {\it GroupLens} Research Group for providing us {\it MovieLens} data and the Netflix Inc. for {\it Netflix} data. This work is partially supported by NSFC (61374177, 71371125 and 71171136), MOE Project of Humanities and Social Science (13YJA630023) and Shanghai First-class Academic Discipline Project (S1201YLXK).

\end{document}